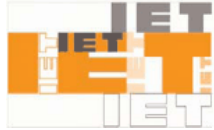

**António Brandão Moniz**
(email: antonio.moniz@kit.edu)

# Robots and humans as co-workers? The human-centred perspective of work with autonomous systems

**IET/CESNOVA**
**Enterprise and Work Innovation pole at FCT-UNL**
**Centro de Estudos em Sociologia**
Faculdade de Ciências e Tecnologia
Universidade Nova de Lisboa
Monte de Caparica
Portugal

# Robots and humans as co-workers? The human-centred perspective of work with autonomous systems

**António B. Moniz**

(Universidade Nova de Lisboa, Portugal and Karlsruhe Institute of Technology, Germany)

## Abstract

## Indice

# Abstract

The design of work organisation systems with automated equipment is facing new challenges and the emergence of new concepts. The social aspects that are related with new concepts on the complex work environments (CWE) are becoming more relevant for that design. The work with autonomous systems implies options in the design of workplaces. Especially that happens in such complex environments. The concepts of "agents", "co-working" or "human-centred technical systems" reveal new dimensions related to human-computer interaction (HCI). With an increase in the number and complexity of those human-technology interfaces, the capacities of human intervention can become limited, originating further problems. The case of robotics is used to exemplify the issues related with automation in working environments and the emergence of new HCI approaches that would include social implications. We conclude that studies on technology assessment of industrial robotics and autonomous agents on manufacturing environment should also focus on the human involvement strategies in organisations. A needed participatory strategy implies a new approach to workplaces design. This means that the research focus must be on the relation between technology and social dimensions not as separate entities, but integrated in the design of an interaction system.

**Key-words**: work organisation, complex work environments, human-computer interaction, manufacturing, technology innovation, autonomous systems, robotics

**JEL codes**: J81, O33

# Introduction

In this working paper the social aspects related with new concepts on the complex work environments (CWE) are analysed, especially those that configure the design of work organisation systems with automated equipment. In such environments the work with autonomous systems represents specific options in the design of workplaces. Technology can be defined not as a product (or an equipment) designed and marketed, but as a social relation that integrates the equipment and working tools, the operators and the material to be transformed. From these relations the concepts of "agents", "co-working" or "human-centred technical systems" reveal new dimensions related to human-computer interaction (HCI). This last concept should not be only defined on the base of configurations of technical systems, but more in terms of organisational configurations, because this concept of organizational configurations under CWE leads to new pattern of human-machine interfaces. The design of work organization implies the definition of tasks to be accomplished by humans with major or less autonomy, using tools and intelligent equipments. It implies also the definition of principles to regulate the means of involvement of humans in the control of those equipments as system agents. With an increase in the number and complexity of those interfaces, the capacities of human intervention can become limited, originating further problems.

Hereby, we will use the case of robotics to exemplify the issues related with the use of automation in working environments and the emergence of new HCI approaches that would include social implications. Some of the most critical approaches on the development of robotics lay on the question if their use may lead to labour displacement or substitution. Following this direction, it would be necessary to know if it can lead to an extension of the digital divide. Would job profiles improve as robots take over dangerous, dull and dirty jobs, as promises for the adoption of these technical changes? And who is the "end-user" of robotics? The answer of these questions marks a debate field of social sciences for several decades. The trends show that the "classical" questions about the social impacts still remain important. But the most recent developments on robotics demonstrate the need to revisit those debated concepts and to increase the collaboration with social scientists among the engineering and computer scientist research teams.

In an interesting article published in 1996, Ejiri approached the future development of robotics. And already there he verified that "many robotics researchers believe that autonomous robots will play an important role in our future society" (Ejiri, 1996: 3). This researcher from Hitachi understood that most problems could be more visible on the possibilities for mobility and motion control, or on energy and battery developments. But a lack of research was done on "machine reliability". He concluded on the state of the art of that time that "we should direct efforts towards providing assistance to human drivers" (idem). And he foresees the development of robotics into this direction when he provides the example of medical field applications: "we have to note that the final goal should not be an automatic surgery machine, but a machine with the capability to help a surgeon as a skilled assistant" (Ejiri, 1996: 4). This is one of the main discussions in this chapter about what would mean the role of agents in the HCI (robots, human operators, other automated machinery, sensors). That means also the discussion about the technological developments in each of the elements that participate in the working environment, and their implications in the way humans work and use such agents or elements of their work environment. This includes the definition of decision process in such CWS. That contributes to consider what an autonomous system in the production sphere is, and what the end-user is with capacity of decision responsibility that can affect safety and quality of work. The answers to that can clarify the role of human work in the increasingly automated spaces.

However, are we talking about a new empirical field? In a certain way, yes. Studies on technology assessment of industrial robotics and autonomous systems on manufacturing environment should also focus on the human involvement strategies in organisations. A needed participatory strategy implies a new approach to workplaces design. Such design implies more interaction with robots, different competences and responsibilities. The research on software development to integrate knowledge based systems into automated and programmable machinery is also an empirical field where the space of tacit knowledge can raise new problems for the formalised knowledge. Several sociological studies were aiming those topics, but almost no research was done in the framework of complex manufacturing systems environments.

For such aim the following research questions can be presented: are there new concepts dealing with the relation of automated systems and job design? Or still the same type of issues are been re-visited in the last years? Which concepts have been accepted and which not? In order to analyse in more detail the concepts on human-machine co-working in this perspective of complex integrated manufacturing systems, this chapter will try to respond to that question. Hereby the guiding hypotheses agree with the conviction that working with autonomous agents is increasing the safety problems and implying a shift in the framework of the relation of humans with their work environments. New questions must be developed to understand newly emerging problems of allocating, monitoring, and diagnosing responsibilities in such systems.

## How to define autonomous systems?

A crucial question at the very beginning is what autonomous systems (AS) are in fact? One can consider as being one network or sets of networks under a single control. This definition considers AS as a feature of internet systems, where one can define algorithmic conditions for autonomous operation. Thus, this definition can be applied either to internet management, or to other machines that work without direct human control.

In a recent report from the US Department of Defence, autonomy is defined as "a capability (or a set of capabilities) that enables a particular action of a system to be automatic or, within programmed boundaries, 'self-governing'" (DSB-DoD, 2012: 10). That does not mean independent machine (computer, robot) decisions and machine uncontrolled action. But technical capacities and features usually lead to organisational models where jobs are displaced with the argument of need to be used on dangerous environments (or military actions).

When one defines that "all autonomous systems are *supervised* by human operators at some level, and autonomous systems' software embodies the designed limits on the actions and decisions delegated to the computer" (idem) that should mean that autonomous systems in the manufacturing industry are mainly supervised by humans. Here we must consider the use of AS not only for "dangerous, dull and dirty jobs", but also for structured environments with predefined and precise task design, as it occurs in production sites or in professional services. Anyway, the meaning of "supervision" in this context must be analysed with further detail.

Another document from the US Department of Defence states that "Dramatic progress in supporting technologies suggests that unprecedented levels of autonomy can be introduced into current and future unmanned systems" (DoD, 2011, p. 43). The definition of autonomy is given as "an automatic system can be described as self-steering or self-regulating and is able to follow an externally given path while compensating for small deviations caused by external disturbances. However, the automatic system is not able to define the path according to some given goal or to choose the goal dictating its path. By contrast, autonomous systems are self-directed toward a goal in that they do not require outside control, but rather are governed by laws and strategies that direct their behaviour" (idem). As this report concludes, an AS system should be:

- self-steering or self-regulating
- self-directed toward a goal

- governed by laws and strategies that direct their behaviour

Is that applicable to all AS? Let us assume that response is positive. If so, and if we consider the use of AS in production areas (for example,), that would mean self-regulating manufacturing systems. Those systems would include autonomous robots, flexible warehouses, production cells, autonomous guided vehicles. These would have the possibility of changing direction according to changing production conditions. Such CWE must be governed by laws and strategies that are directing the AS behaviour. This means highly complex models of work organisation where the competences of system operators should be considered carefully.

For AS, human operators and software developers must create, develop and test control algorithms. However, such systems can include at least basic machine learning procedures. In this way, the autonomous systems can develop modified strategies for themselves by which they select their behaviour or reaction modes without the interference of human operators. The most advanced automated equipment in the production field is able actually to adopt such procedures. But the potential for that development is still very limited. The most interesting results would come with the design of models where human operators can be integrated in advanced HCI systems.

An AS, also in manufacturing industry, can be self-directed by choosing the behaviour it follows to reach a human-directed goal. This happens especially with the recent developments in robotics. As for military or for industrial applications, various levels of autonomy in any system guide how much and how often humans need to interact or intervene with the autonomous system. The human-robot interaction (HRI) approach integrates such understanding. In some other cases, autonomous systems may even optimize behaviour in a goal-directed manner in unforeseen situations (i.e., in a given situation, the autonomous system finds the optimal solution), for example, for inspections purposes. Applications on professional services, like mining, or agriculture and forestry, process industry (oil, chemical, paper pulp), medicine, planetary, or manufacturing application for non-structured production are adequate examples of use of AS. Such applications in CWE can help to respond quickly to unexpected problems and to increment safer conditions for human work, with increase of productive performance. However, as Weiss and colleagues mention that although people are receptive toward working with robots, they would prefer to work alongside a tele-operated robot rather than an autonomous one (humanoid). The condition would be "as long as there is a clear distinction between a human and a robot, in terms of tasks and working procedures" (Weiss et al., 2009). Thus, autonomous agents still have problems in terms of acceptance in work spheres, and that must be taken into consideration [1]. Heyer also concludes that "how robots fit into organisational structures, and how they are accountable to the organisation in terms of safe and reliable operation is yet to be determined" (Heyer, 2010: 4753). In the same direction Groom and colleagues refer that for robot operators the "perceptions of the humanoid as an independent identity would produce distrust and disliking of the robot" (Groom et al., 2009: 32).

Under this possibility, we will use the concept of AS applied to robots and other intelligent systems in the production areas that are able to autonomously operate in complex environments. Industrial production presents complex environment while the operational tasks can vary in higher degree in small batches, which requires an increased flexibility. Additionally, the adoption of just-in-time (JIT) strategies [2] increased also the complexity and variety of tasks per considered unit. In the same production line, the varieties of components are able to increase and are depend fully on the demand side. The successive implementation of new logistic policies at the shop floor introduced also an increased variety of machinery interaction and mobility of parts using different support infrastructures. This means the introduction of new ICT capacities in the shop floor environment, with high level of information processing capacities for a wide variety of sensors, programming elements, integration of communication technologies (RFID, laser or others). The new elements of AS are the following:

- Perception,

- Control,

- Learning, and

- Systems.

---

[1] As Yee, Bailenson, and Rickertsen refer "while most studies have found that interface agents have positive effects on task performance, these effects are overall actually quite small" (Yee, Bailenson, and Rickertsen, 2007: 6).

The perception includes the means through which AS can interact with the environment and with humans. That includes sensing features, the representations of information, the modelling and management of uncertainty, data fusion and perceptual interpretation. Most of these elements are usually software-based, but most can integrate information provided by human operators. The representations of information are certainly pre-defined by humans according to the working task objectives. They must be clearly understood, and the trend is increase the intuitivity of such representations. The limits of uncertainty are fixed by humans and the conditions to be managed can also be defined by humans. Finally, the perceptual interpretation can be done according to the frames established by human operators.

The control of individual machines can be done by human operators, but gradually is possible to leave it to AS to a certain level. The complexity of AS enables also the possibility to control heterogeneous groups of platforms and sensors. This can be found as positive the decrease the stress level of task completion, and to help the human operator to concentrate on more complex tasks. AS are also able to contact and interact with the environment and each other. This is also an important condition to define the level of autonomy. Some IR integrated in flexible manufacturing systems (FMS) can control other equipment and interact with the environment.

The supervised and unsupervised learning in unstructured and dynamic environments is also an important element of AS. In some cases human operators can use the collected information as an input for new machine learning process. In other cases, multi-agent learning is possible, and pattern recognition and concept formation. In all of these approaches the involvement of human operator knowledge is very important to the continuous capacity of AS operation.

The design and optimisation of "systems of systems" is becoming a crucial element of definition of work organisation model. That means a high capacity to involve operators in the supervision of decision process with the help of AS. The modelling and management of complexity is becoming a task that can be allocated to automated systems, and a large scale systems theory can be elaborated in the frame of an organisational strategy for a CWS. Hereby, the modelling of information flow constitutes also an element of AS.

---

[2] This rationalization process in industry was developed in the 1990s in a global level, and introduced additional needs in terms of quick responses to client and supplier demands from main manufacturing companies. JIT strategies also introduced new elements of space management to the production networks, where the information and communication technologies played an important role to enable the installed capacities. Automation and autonomous systems could contribute to enable those purposes of rational and complex production strategies.

Taking the technological options into account, do we understand the industrial (IR) and service (SR) robotics as autonomous systems? Obviously yes, because they are "acting" without direct control, although operating in complex industrial environments. The complexity of manufacturing equipment is revealed, in particular, by the need of programming or re-programming (off- or on-line), and by their increased multi-functionality. For instance, new programming methods can enable the automation of small lot sizes or even single work pieces, hand drawings made with a digital pen can be transferred into robot programs automatically, or robot trajectories can be defined by guiding the robot using tactile feedback. Such flexibility can improve the task performance, with direct effects on quality, safety and productivity.

To understand how far these systems still are very difficult to achieve, one can retrieve some comments from the above-mentioned report of the US Department of Defence, where they mention that "the fact that autonomy software interacts with a dynamic environment in a non-deterministic manner is particularly challenging, especially for agencies that are used to full-path regression testing that validates every individual requirement". In other words, robot developers could not achieve the capacity to enable HRI in non-deterministic ways in dynamic environment. Under such conditions, human operators can achieve these capacities without major problems. One can say that is a human feature. And to develop those features on machines, the complexity is so great that the risk of trying it is not worth. At least, in the present time.

And the mentioned US DoD report add that "the reference framework highlights the need to confirm how the autonomous system provides its operator and upper echelons of direct and indirect users with the basis for making the decisions delegated to it during different mission phases [here one can replace the word 'mission' by other non-military reference-abm]. It also highlights the need for measures and models of the dimensions of system resilience/brittleness that can be used early in systems development as well as later in T&E [test & evaluation-abm]" (p. 11). The problem remains on the capacity for allowance of system autonomy. The definition of a decision process is critical to the capacity of that autonomy. In military action the commando structure evidences the simple decision process. But in "civil" action, in real economic structures like companies and other organisations, the definitions of operators and users lead to very diverse possibilities of decisions. That is due not only to hierarchical positions in the decision process, but more likely to the level of professional competences involved in that process. And the competence is related to the training received (formal or informal), and also to the experience to process the working tasks. The "system resilience" must also integrate the environmental dynamics that can be complex in military or civil conditions. Thus, the dimensions of that resilience have to be included in the system development. And that task is usually allocated to the social sciences approach.

As Drury and colleagues notes that the greater or the lesser amounts of HRI awareness are needed depends upon the level of autonomy that the robot achieves. Therefore, "the expectations of awareness need to be tailored for the anticipated level of autonomy. The HRI awareness base case can be generalized to cover multiple humans and robots coordinating in real time on a task" (Drury, Scholtz and Yanco, 2004: 2).

## The increasing relevance of industrial robotics in services economies and it's labour impact

Industrial robotics (IR) can be understood after a decade of the 21$^{st}$ century as production equipment which relevance is decreasing in the more developed and service-oriented economies. With an increased process of economic globalisation, the manufacturing industries have been transferred to emerging economies (China, India, Latin America and other economies) and with this de-localisation process also implies a technology transfer and de-localisation of company sites into such new emerging markets (Moniz & Paulos, 2008; Paulos & Moniz, 2008). The use of robotics in industry can also follow this process. This de-localisation process also may have implications on the labour market. But does this empirical evidence means also the decrease of the number of IR installations in developed countries? The answer apparently is no, according with the data presented by the International Federation of Robotics – IFR (IFR, 2001). But that seems to be contradictory: the national economies that are more dependent on service sector income are those that have a growth in the number of IR? Here the answer is yes. These economies are using more intensively IR while the levels of automation increased with the transfer of manual labour to other industrialised economies and with the decrease of jobs in the manufacturing sector in the service-oriented economies where IR are installed.

As Morioka and Sakakibara (2010) writes that "to maintain the competitiveness of manufacturing in those countries where wages are high, it would be advantageous if improved efficiency and cost reduction could be achieved by means of automation centred on industrial robots, with additional tasks such as parts feeding being performed by human". As mentioned by Decker and colleagues (2011) "the major distinction between service and industry robots is based on the characteristics of services: they are immaterial and thus experience goods whose quality can only be assessed once they are actually used by the customer(s). The simultaneity of production and consumption as well as the consequential direct relation between service provider and customer is the reason why services cannot be stored, exchanged, or sold again" (Decker et al., 2011: 39). Therefore, from the point of view of work performance, manufacturing production or production of services is similar. They are both related to a value added to a (material or immaterial) product. Automation can increase the productivity adding more product units for the same amount of time units, or used equipment or even direct involved people in that production process.

The extensive introduction of robotics in manufacturing industry has been a fact in the last decades. For the major sectors, and in the cases where quality control is a critical feature of production process, automated systems are being introduced. Robots can also be effective in areas where there are skill shortages. In particular, they are being used in those repetitive and with worse working conditions, leaving to the human workers the highly qualified tasks with increased involvement in the job decision process. At least that has been the argument for the robot manufacturers, but that can be not always true [3]. The rationalisation process in the production sectors is pushing organisations to introduce new automated systems, and that would also mean an increase of robotics use in most manufacturing sectors. Automation increases the economic performance of those sectors but also the skills needs for operators of AS (including robotics). Thus the main motivation is not necessarily a social one (to increase safety and better working conditions) but to respond a management strategy (rationalisation) that requires an increased investment on automation. And this management choice induces an increase of competence requirement and skill needs.

The decrease of annual supply for Asia was felt in different moments in the last decades, as 1998 and 2001, but overall the number of new IR installations is much higher than in the other regions of the world. It is worth to underline the increasing number of IR supply in Europe, and a recent strong growth in America. A very high number of IR is to be supplied and installed in Asia according to IFR (2012). That can happen in America too, although in a different rate [4]. Europe can experience fluctuations in this statistical indicator [5]. Moreover, the expectation for 2014 point out an unprecedented higher number of IR installations in all of these regions [6]. In 2011, Japan was again the biggest robot market in the world, which is almost a constant in the history of IR supplies. The robot installations in Japan continued to recover after the 2011 disaster and increased by 27% to almost 28,000 units in relation to 2010. The automotive industry and most of all other sectors were increasing robot investments above average [7]. The Chinese market is acknowledging a recent growth where is estimated to have in 2010 52.3 thousand operational stock and 191.3 thousand in 2015.

If we consider the sectors with new IR supply at the world level two different periods can be visible: one in 2004-05, and the other in most recent years (from 2008 until 2010). In the first one, there is a constant increase in the number of IR supplies in almost all sectors. A special relevance can be made for the electrical industries that almost achieve the same number of yearly supply of the motor vehicle industry, or the automotive components. In the first case, it was noticed a clear increase in the number of IR for the electrical industries, and also for the automobile industry (motor and automotive parts) and chemical. This trend is much more intense in the following years.

The electric and electronic sector (including computers) became the one where most IR has been installed. With the exception of 2009 that was a reflex of the 2008 financial crisis around the world, no other sector experiences such expressive change. As mentioned in a recent report from IFR (2012), they mention that the "sales to all industries except for automotive and electrical/electronics [sector] increased by 37% in 2011. The robot suppliers reported a considerable increase of customers in the past years. However, the number of units which is ordered by these customers is often very small". That can mean as well that all sectors of manufacturing industry are becoming more used to automated equipment, and especially to robotics. That is not anymore a specificity of the automotive sector and/or the electronics.

[3] As the Gorle and Clive report for IFR underlines, "electronics or automotive are high users of robots and without them parts of the industry could not survive even in the low cost countries. It also varies between countries, with higher proportions of robot dependent jobs in high cost countries such as Japan and Germany, and lower proportions in low cost countries such as China and India" (Gorle and Clive, 2011: 5).

[4] In North American countries the operational stock of multipurpose industrial robots is acknowledging a fast growth (from 123.7 thousand in 2004 up to 248.2 in 2015).

[5] Europe as whole is also witnessing a constant growth in the number of operational stock (from 279 thousand in 2004 up to 422.5 in 2015, with special relevance for Germany with 176.8 thousand to be expected in 2015).

[6] The IFR estimates that the total worldwide stock of operational industrial robots at the end of 2010 was in the range of 1,035,000 and 1,300,000 units. In 2006 the world total stock of operational industrial robots was 951,000 units. The recent trends in terms of investment on new IR in the recent years can notice a slight variation, but overall there is an increase on the worldwide supply of IR. There are two moments of significative decrease due to the global financial crises, but the recovery was felt almost immediately.

[7] According to IFR again, Japan had in 2010 around 308 thousand operational stock, and the next year 307.2 thousand. The expectations for 2012 are less than 300 thousand, and for 2013 up to 306.5 thousand, but still not achieving the same amount as in 2009.

Finally, the measure of robot density is the number of multipurpose industrial robots per 10,000 persons employed in manufacturing industry [8]. This "density" is usually higher in the automotive industry than in the "general industry" (which is all industries excluding the automotive industry) [9]. The estimated average robot density in the world is about 55 industrial robots in operation per 10,000 employees in manufacturing industry. This gives a real idea of the level of investment made by three countries (Korean, Japan and Germany, and more recently, China) with a substantial increase in the robot density of the automotive industry in the last decades.

The considerable high rate of automation of the automotive industry compared to all other sectors is demonstrated in the evaluation of the number of industrial robots in operation per 10,000 employees in automotive industry and in all other industries. The technological level with respect to robotics is rather homogeneous in the motor vehicle industry (1 robot per 10 workers). The automotive industry will continue to be the innovator for new technology, as IFR states in their 2012 report [10]. Already in the previous year IFR mentions that "the overall conclusions indicate that in almost all the surveyed countries, not only the potential for robot installations in the non-automotive industries is still tremendous, but it is also considerably high in the automotive industry among the emerging markets and in some traditional markets as well. This is mostly due to the necessary modernization and retooling that is needed in these markets" (IFR, 2011: x).

In the same report IPR states that "the main impulses are coming from North America, China and other South-east Asian countries. Investments in Japan will gain momentum as reconstruction and new projects are carried out in the coming months. Japan is likely to return at the top of the robot market in 2011. As a consequence of the disaster in Japan, Japanese companies have been trying to diversify their production geographically. This will result in considerable investments in robot installations in Asian markets as well as in Europe and in North America. The robot supply to the Republic of Korea will only slightly increase after the huge investments in 2010. Robot supply to China will surge and finally at least in 2014 China will be on top of the robot markets. The robot sales in Europe will increase below average because of a rather moderate increase in investment by the western European countries. The robot installations in the eastern and central European countries will surge in 2011. However, it is still possible that due to a shortage of components and capacity problems a part of these expected robot installations will have to be shifted to 2012. The automotive industry is continuing to be the main driver of the growth in worldwide robot installations with investments in new technologies, further capacities and renovation of production sites" (IPR, 2011: xi). This would have impacts in terms of labour market of skilled labour that operate IR and other AS. It can be expected to increase the number of such systems in sectors where costs can't be downgraded (for example, automotive and electronics). This can produce a evident "digital divide" in the sense the highly qualified jobs (for example, IR operators and technicians) will remain in countries with higher labour costs, and lower qualified jobs will be allocated in industries de-localised to lower labour cost countries.

---

[8] In fact, one can consider this group of countries (Japan, the Republic of Korea and Germany) as the most automated countries in the world . Already in 2005, 352 robots per 10,000 persons employed in the manufacturing industry were in operation in Japan and 173 robots in the Rep. of Korea. With 171 robots per 10,000 employed in the manufacturing industry Germany was already the country with highest robot density in Europe. This means also that Japan decreased this figure of density and Korea and Germany increased their density.

[9] In 2011, the Republic of Korea reached the highest robot density in the world with 347 industrial robots in operation per 10,000 employees. The reason is the continued large volume of robot installations in the recent years. The robot density in Japan increased to 339, that of Germany to 261.

[10] The Gorle and Clive report refer that "if a car plant has 500 robots this could require say 50 skilled technicians for the robots" (Gorle and Clive, 2011: 23). As the automotive industry has 365 thousand robots according to IFR that would mean around 36 thousand dedicated staff.

For the service robots (SR) for professional use, 39,900 units were installed up to the end of 2006 [11]. However, the total number of professional service robots sold in 2011 rose by 9% compared to 2010 to 16,408 units up from 15,027 in 2010 [12]. Application areas with strong growth were defence, rescue and security applications, laboratory robots, professional cleaning robots, medical robots and mobile robot platforms for multiple uses. The underwater systems were expected to decline the growth rate. In the more recent years, become clear that the military and defence sector was the one where most of the professional service applications have been used (around 6 thousand new applications per year). The "field sector" (agriculture, forestry, milking) has a higher growth rate [13]. In 2010 the sales of medical robots increased by 14% compared to 2009, accounting for a share of 7% of the total sales of professional service robots. The most important applications are robot assisted surgery and therapy. The logistic robots are in the 4th place in terms of professional service applications, which are courier and mail systems as well as automated guided vehicles (AGV) for manufacturing factories. The average service life might in fact be as long as 15 years, which would result in a worldwide stock of 1,300,000 units in manufacturing industry (according to IFR, 2012) [14].

Around 4 million workers are dealing directly with robots in their job tasks with an average of 3 workers/robot in a 1.3 million robot population (see also Gorle and Clive, 2011). Some new roadmaps on robotics also underline that a more widespread use of robots may lead to further labour displacement and an extension of the digital divide (see EUROP, 2009). This contradicts the assumptions of IFR about the positive effects. Anyway, the increased application of this equipment for production activities may lead to the exclusion of parts of the society from the benefits of advanced robotics. On the other hand, job profiles can improve as robots take over dangerous, dull and dirty jobs not only in the manufacturing industries. The real labour market impact should be analysed more carefully and there is not enough empirical evidence of the impact on employment due to the diffusion of AS.

At the same time, there are no data on the level of „intelligence" and „autonomy" of these equipment's. One can still question about how far important are the robotic equipment (IR and SR) in terms of influence in the production sphere, of market dimension and share [15]. If one considers the average robot unit price and the possibility to be installed in small- and medium-sized companies (SME), a whole new market is increasing, and its social implications can represent new trends. In fact the average price fell to about one third of its equivalent price in 1990, which means that automation is becoming economically more affordable. At the same time, the robot performance such as speed, load capacity, mean-time-between-failure has dramatically improved. These improvements provide a faster return on investment, particularly for small, short-run batch production.

---

[11] Projections made then for the period 2007-2010 was about 35,500 new installations for that period. Turning to the projections made in 2006 for the period 2006-2009, the stock of service robots for professional use was forecast to increase by some 34 thousand units.

[12] As the recent IFR report states, "since 1998, a total of more than 110,000 service robots for professional use have been counted in these statistics. It is not possible to estimate how many of these robots are still in operation because of the diversity of these products. Some e.g. underwater robots might be more than 20 years in operation. Others like defence robots, may only be for a short time in operation" (IFR, 2012: 15).

[13] As IFR states in the 2012 report, "turning to the projections for the period 2011-2014, sales of professional service robots are forecast to increase by about 87,500 units. Thereof, more than 25,500 milking robots will be sold in the period 2011-2014. They are followed by service robots for defence applications with more than 22,600 units. This is probably a rather conservative estimate. These two service robot group make up 55% of the total forecast of service robots" (IFR, 2012: xv). But other service sectors are also acknowledging recent changes.

[14] In the professional service sectors the figure is not clear. If it was about 31,600 units installed up to the end of 2005 and 34 thousand new service robots for professional use to be installed for the period 2006-2009 (IFR, 2006: ix), and the total number of this type of service robots that were sold in 2010 rose to 13,741 units per year (IFR, 2012: xiv), one can have an estimate of around 80 thousand stock units.

[15] In terms of market value of the professional service robots is around US$3.2 billion (value of new sales), and US$538 million for the service robots for personal and domestic use (figures for 2010). The sales of IR in 2010 reached the value of US$ 5.7 billion. The worldwide market value for robot systems in 2010 is therefore estimated to be US$17.5 billion in 2011 (IFR, 2012: x).

The impact in terms of employment was recently presented in an IFR report (Gorle and Clive, 2011). The robot industry and operation generated 300 thousand jobs worldwide according to this report. It focuses also the attention to sectors where the need for reduction of production is becoming more important. One of the sectors is mining and another is the electric storage media. In both cases the production growth will be evident in the next years, and when the breakthrough occurs, the employment connect with the application of robots and autonomous systems in such sectors will increase

## The involvement of workers with autonomous systems

The concept of "Anthropocentric Production System", developed in the decade of 1990 in the framework of European projects [16], is a coherent set of technological and organizational innovations to improve productivity, quality and flexibility. "The production system that fits this condition is a computer-aided production system strongly based on skilled work and human decision-making combined with leading edge technology. It can be called 'an anthropocentric production system" (Lehner, 1992). This concept corresponded also to new developments in the autonomous systems research. In early 1980 decade "most industrial robots have neither tactile sensing nor vision and thus cannot operate effectively outside a rigidly controlled environment" (Medeiros and Sadowski, 1983: 3). However, in present times, intelligent and autonomous systems are able to operate in complex environments and with complex senses and reasoning principles. These systems should operate jointly with humans and particularly with workers at the shop floor of industrial manufactures. Thus, particular attention must be shifted into the competences management and skills needs for such involvement.

It is common to say that IR performs pre-programmed actions in a specifically prepared and highly structured environment. Such environments can be found in manufacturing industry but usually in special working areas or cells. In such cases they have no need for perception and on-the-task human interaction. With the increased involvement of humans and autonomous systems, or co-working scenarios, the robots and humans must cooperate to fulfil a common goal. In fact, in that scenario the environment tends to be less structured and the human operator (or the direct workers) is supposed to "program" the robot as the work unfolds.

---

[16] The most well-known project was the ESPRIT 1199 on Human-centred CIM systems. However, other projects and networks also referred to this concept. More information on this see Moniz (2012).

Some robot manufacturers (FANUC, for example) developed a new cell production assembly system, in which physical and information supports are provided to the human operators, reinforcing the safe cooperation with the robot while providing instructions on the operations to be performed. The safety management for the cooperation is new from the viewpoint that industrial robots can cooperate with human operators in automatic operation without stopping, which can achieve higher productivity (Morioka and Sakakibara, 2010: 12). In such situations flexible automation strategies can support the human work and a distributed decision making. A decentralized organization of work, with flat hierarchies and a strong delegation of power and responsibilities, especially at shop-floor level would be the more suitable model to integrate such automated systems. This can be especially important while "human interaction aspects, especially uncertainties caused by human behaviour, and the requirement for parallel task execution in real-time have not been sufficiently researched" (Kobayashi et al., 2012: 3)

This organisational strategy implies a reduced division of labour, and a managerial principle of continuous and product-oriented up-skilling of people at work. Such principles and strategic orientation is not very easy to find in the manufacturing industries. The aim for product-oriented integration within the broader production processes can imply a more flexible and human-centred organisational strategy, but SME cases can be found easier, than in larger corporations.

The research activities at the European level on the concept of new working environments gave considerable attention to the challenges of the increased competencies of people working together with automated technologies. In fact, the European Commission (EC) coordinated research activities during the 80s in the field of Anthropocentric Robotic Systems that influenced the ESPRIT program during several decades, and a wide group of European social scientist. The attention to such field does not come only after 2000 with the so-called "Lisbon Strategy" but from decades earlier, for example, with the activities at the Forecasting and Assessment in Science and Technologies (FAST) unit of DG Research. This EC unit paved the ground for new networks and research projects [17].

In order to improve the productivity and to use the technical capacities of new robotic systems, it is always needed to develop further involvement of operators in the decision process. But such involvement needs a participatory strategy for workplaces design, and that is not always available. Thus the limit to this development starts at this point. The organisation design is a prior element to the technological development and innovation.

Moreover, such organisational design implies more interaction with robots, as is the case for the manufacturing industry and for some professional services that need robot applications. In this last situation are the examples for medical and chirurgical robots, field robots, logistics and others. As presented in previous sub-chapter, this equipment market is much larger than the one related with other personal and leisure robotics.

The need to relate the working perception with AS (e.g. autonomous robotics) is still a research topic that needs further evidence from the sociological perspective. For example, the cognitive task automation may lead to over trust, complacency and loss of the necessary work environment situation awareness. This can lead to safety problems within CWE using such technologies. This can be also a major constraint in the organization of working teams under CWE. It may end up into an operational gap between system developments and its understanding and usability, by operators. In fact, many concepts issued from the work organization analysis, are connected with other concepts such as motivation, alienation, satisfaction, productivity, innovation, flexibility and business processes, learning organizations, networks and virtual enterprises. And these concepts must be taken in consideration to foster the analytical dimension on the use of AS under participative structures of industrial organisations.

---

[17] More recently, the EUROP platform - or the European Robotics Technology Platform -, was founded in 2005 and is an industry-driven framework for the main stakeholders in robotics to strengthen Europe's competitiveness in robotic R&D, as well as global markets, and to improve quality of life (http://www.robotics-platform.eu/cms/index.php). The platform established a Strategic Research Agenda for robotics (SRA) where it prioritised the research and development efforts in Europe and encouraged cross-fertilisation among all robotics domains and beyond.

As Thrun underlines, "human-robot interaction cannot be studied without consideration of a robot's degree of autonomy, because it is a determining factor with regards to the tasks a robot can perform, and the level at which the interaction takes place" (Thrun, 2004: 14). The image one retains is that the environment of industrial settings is quite structured and engineered, and the amount of required autonomy should not be very high. Nevertheless, robotics research has been focused on autonomy enabling. The reason is based on the need to operate closely to humans in a safety concerns. Other robotics researchers say that "in order to guarantee a safe human-robot interaction in the shop floor environment, there must be a continuous and intelligent monitoring process that can provide the system with essential information regarding users' localization as well as hazardous areas. With an increasing number of accidents resulting from overlooking or overriding traditional (reactive) safety devices there is a clear gap in advanced proactive unobtrusive safety tools (Ribeiro, Barata and Barreira, 2009: 31). Thus, we may conclude that HRI should consider not only the robot's degree of autonomy, but the organisational flexibility (with workers participation strategies in the decision process) where such robots are introduced. And, unfortunately, the system designers don't take into consideration such anthropocentric approach.

As Lenz refers in his thesis, "the collaboration between human and robot has been announced as a promising approach to solve these challenges, because it teams the strength and the efficiency of robots with the high degree of dexterity and the cognitive capabilities of humans into a flexible overall system. As consequence of current flexible automation techniques including flexible manufacturing systems (FMS) and reconfigurable manufacturing systems (RMS), a recent trend in robotics focuses on new generations of robots with the capability to directly assist humans. This bridges the gap between fully automated systems and fully manual workstations" (Lenz, 2011: 2). And also he concludes that "the advantage of the potentials for humans and robots to work together as a team is only in early stages and needs a safe, robust and efficient realization" (idem: 3). This represents the main problem for a co-working system in any robot application where the interaction with humans is needed. Normally, it is assumed that the assembly plans are known to both human and robot a-priori and that the attention of the human lays on the joint assembly task to provide balanced action coordination. As Lenz points out, under these assumptions, some mechanisms as "task sharing" and "joint attention" are disregarded. This can be a problem while such mechanisms are fundamental to achieve a safer co-working action.

The HRI systems should be designed support tacit or formal knowledge in the production process. In manufacturing industry, the integrative tasks in advanced automated systems can be taken by human workers. The same applies to the control tasks. Humans are also better at dealing with unexpected events to keep production lines running. Interaction of humans with robots increases the importance of such aspects. Intuitive programming, augmented reality and programming by demonstration are interesting concepts that deal directly with safety, control and participation in the decision process.

While most robots operate in industrial settings where they perform different tasks (assembly, welding, painting, drilling, etc.) the direct interaction implies basically a risk assessment in terms of safety[18]. This refers not only to the ergonomic dimension, but it clearly strengthens organisational issues (social implications) where different options are available. "Existing ISO robot standards have been developed with a limiting focus on industrial use, because robots have formerly only been considered as valuable tools for manufacturing in industrial environments. Therefore, these norms are mostly only applicable to static industrial tasks such as lifting heavy parts, machining various metal and non-metal components, and joining large panels. Such applications demand the use of robots with large and powerful machines, which result in highly hazardous collaboration partners. Such risks for the human are avoided by separating the workspace of robots with real or virtual cages or in time (Lenz, 2011: 15).

There is a need to widening the perspective with respect to the social implications within the intuitive interaction between humans and robot systems. Hereby the different technical options of intuitive interaction can be analysed and assessed with regard to increasing decisional options for the human operators. The new ISO 10218-2 norm sets the requirements for the layout design of the workspace around the robot, including safeguarded spaces (where humans are separated from the robot and protected by safeguards) and collaborative spaces where humans are not separated from the robot and hence the robot shall apply the control limits mentioned above [19]. The switching between autonomous and collaboration mode needs to be performed in a way, that no person is endangered. It specifies also the operating modes that must be designed into the robot's control function when collaborating with a human in the collaborative workspace.

There are however fields of required technical improvements, as the choreographing the task movements, for instance loops and conditions, without requiring extensive programming competencies. "The observation of human actions by an assistive robotic system can, for example, be used in an assembly task to act pro-actively with the preparation of future steps based on the current action of the human. (…)The experiments showed that with the right abstraction level regarding sensor information, the actions can be recognized by means of hand velocity, acceleration, and jerk. This enables the transfer of the recognition models to the human robot collaboration scenario" (Lenz, 2011: 5). Specification of how external sensing should be used for new types of motions or for handling unknown variation is still needed. Therefore, the collaboration features still must be defined. Autonomous systems still present intelligent limitations to be able to be integrated in human-like work environment, except in the most repetitive and simple tasks that are usually applied to mass production.

At the same time, another problem is not yet solved in order to establish the conditions for co-working: one cannot instruct a robot in the same way that one would instruct a human worker how to carry out a task. Thus, the robot "training" must be done differently and the integration for similar tasks is not yet achieved.

---

[18] see EN ISO 10218-1:2006, *Robots for Industrial Environments – Safety Requirements – Part 1: Robot*, 2006
[19] The DLR lightweight arm is used to build a sensor-based robotic co-worker for a safe and close cooperation and presents strategies for safe interaction with the human (Haddadin et al., 2011: 12)

It is desirable that for the robot operation it could be possible to manually guide a robot to the positions of interest, or even along the desired paths or trajectories if human accuracy is enough. This would need always an increased interaction with the human operator. Also, it should have more simple ways to make use of CAD data whenever available. There are new possibilities for using different complementary modalities (paths of communication between the human and the robot), such as speech and gestures. These features are increasing the HRI developments and clearer definition of co-working activities.

Thus, new ideas are to be developed to understand newly emerging problems of allocating, monitoring, and diagnosing responsibilities. During the technical design, the arrangement of the man-machine interface and the design of the control program are of great importance regarding the decision authority. Therefore, in order to allow humans to take the responsibility for functioning robots, these must be controllable in the sense of transparency, forecast and influence. But which humans can take that responsibility? Are there the operators? Or just the technicians? The answers to such questions are not easy to be responded.

According to the EUROP vision, "in the short term robots and humans will work beside each other and, in some cases, interact directly. In the mid-term robots and humans will cooperate and share space with each other, both at work and at home. Robots will perform more complex tasks without constant supervision. Only in the long term will humans and robots become more integrated and will the sophistication of the interaction increase" (EUROP 2009). Such vision encapsulates several common sense ideas and some wide spread concepts. However they can be discussed and much research is needed to improve the knowledge around such principles.

As the increased use of robot installations in working environments increases, the need for interaction pushed also more research on software development to integrate knowledge based systems into automated and programmable machinery. As Kochan concludes in his work, with the new developments of robotic manufacturer in the direction of safe robot, it seems to be clear, that human operators and robots will soon be working together in one workspace without the need for the robotic system to suspending its work when humans come too close to the robot (Kochan, 2006: 13).

A parallel issue is related with the decision-making by mixed human-machine teams. It raises, at the same time, the issue of ethical and legal responsibility. The main reason is because wrong decisions may involve the contribution of machines in various perception or reasoning stages.

## Conclusive remarks

We have analysed some of the social aspects related with the concept of CWE, especially those that configure the design of work organisation systems with automated equipment revealing new dimensions related to HCI. The study of the dissemination of AS in production activities (mostly, IR) led us to understand possible implications to the labour market. In fact, the volume of introduction of new robots in manufacturing industry is much higher than with SR, or even with professional service robots. However expectations point out to a clear increase of this market. Automotive industry sector is the one where most IR are introduced, and the one with highest density. Thus, the type of AS used with interaction with humans can indicate what are the social needs associated to the design of technology. For that reason we pointed out in the introduction that studies on technology assessment of IR and AS on manufacturing environment should also focus on the human involvement strategies in organisations. A needed participatory strategy implies a new approach to workplaces design.

On one hand, does the technology design have consideration for organisational and social dimensions? It seems that the major IR manufacturers do not consider those dimensions. This can lead to further problems on systems implementations in CWE. The workplace design under such environments implies more interaction with robots, different competences and responsibilities.

On another had, is not clear which kind of discussion framework is the social science dealing with under this relationship. The concept of "new working environments" gave considerable attention to the challenges of the increased competencies of people working together with automated technologies. But in the last years social sciences did not produce further knowledge on such issues. Nevertheless, non-technological dimension (sociological, psychological, cultural, and ethical) of technology design should be recognised and taken into consideration.

In such conditions, it can be questionable how far is possible to implement real inter-active procedures. The same would be applied to the use of HRI integrating organisational dimensions. One cannot really speak about "common aims" in co-working environments integrating humans and AS. When one takes workgroup strategies the concept of "common aim" must be taken. Thus how could it be possible to design co-working environments without workgroup strategies? We must conclude that would not be rationally possible.

Today one can understand the wider use of the anthropocentrism concept applied to the production architectures, although intrinsic difficulties can be evident. These difficulties rely either on the side of organizational design (that include co-working features) or on the side of technical development. It is emerging, however, a new indication of the value of intuitive capacities and human knowledge in the optimization and flexibilization of the manufacturing processes. These dimensions were not usually considered. But when there are new risk situations that occur with the use of autonomous systems (especially, IR and SR), those can be elements to consider in the design process. It is becoming evident that is necessary to take into consideration qualitative variables in the definition and design of robotic (IR/SR) systems, jobs and production systems.

It must be already possible to implement knowledge sharing at the workplace. But that is not always recognised when applied to AS, or at least to IR in manufacturing environments. An improved "intelligent" workplace should mean not only an increased capacity of the manufacturing system (that would include robotics, numerical control machine tools, logistics and complex work flows, etc.) in terms of programming, system control or environment data processing. It should also mean the involvement of operators that intervene in the different manufacturing phases. They should become more "system operators", and less "machine operators". The issue of responsibility in the decision process is still not clear: in increased self-controlled system who takes responsibility for unexpected events? Are autonomous systems co-workers of organisational managers? Will it be possible for autonomous agents to achieve tacit knowledge? The answer to such questions need further research evidences.

To summarise we can say that the status of the scientific research on these issues is no longer focused on the human aspects of the manufacturing automation concepts. The focus has been taken on the human-machine interfaces and on the self-governance of autonomous systems (AS). In other words, the focus is on the relation between technology and social dimensions not as separate entities, but integrated in the design of an interaction system.

**References**

Bernstein, Crowley and Nourbakhsh (2007). Working with a robot: Exploring relationship potential in human–robot systems, *Interaction Studies* 8:3, 465–482

DoD - Department of Defence (2012), *Unmanned Systems Integrated Roadmap FY2011-2036*, Reference Number: 11-S-3613, Washington.

DSB-DoD - Defence Science Board (DSB), Department of Defence (2012), *Report of the DSB Task Force on the Role of Autonomy in DoD Systems*, Washington.

Drury, J.; Scholtz, J. and Yanco, H. (2004). *Applying CSCW and HCI Techniques to Human-Robot Interaction*, report (Case #04-0166), Bedford MA, MITRE.

Ejiri, M. (1996). Towards meaningful robotics for the future: Are we headed in the right direction?, *Robotics and Autonomous Systems,* 18 (1), pp. 1-5

EUROP (2009). Robotic Visions to 2020 and beyond – The Strategic Research Agenda for robotics in Europe, 07/2009

Gorle, P. and Clive, A. (2011), *Positive impact of industrial robots on employment*, London, IFR/Metra Martech, 66 pp.

Groom, I Am My Robot: The Impact of Robot-building and Robot Form on Operators, in ACM: Proceedings of HRI'09, March 11–13, 2009, La Jolla, Ca. ACM.

Heyer, C. (2010), Human-robot interaction and future industrial robotics applications, in 2010 IEEE/RSJ: *Proceedings of the International Conference on Intelligent Robots and Systems*, Taipei, IEEE, pp. 4749-4754.

IFR-International Federation of Robotics (2011), *World Robotics 2011*, IFR
IFR-International Federation of Robotics (2012), *World Robotics 2012*, IFR

Haddadin, S. et al. (2011). Towards The Robotic Co-Worker. In C. Pradalier, R. Siegwart, and G. Hirzinger (Eds), *Robotics Research*. Springer, Berlin/Heidelberg, 2011, pp. 261–282.

Kobayashi, Y. et al. (2012): Multi-tasking arbitration and behaviour design for human-interactive robots, *International Journal of Systems Science*, DOI:10.1080/00207721.2011.625477

Kochan, A. (2006). Robots and operators work hand in hand. *Industrial Robot: An International Journal*, 33(6): 422–424.

Lenz, Claus (2011). *Context-aware human-robot collaboration as a basis for future cognitive factories*, PhD thesis, Technische Universität München, 145 pp.

Medeiros, D. and Sadowski, R. (1983). Simulation of robotic manufacturing cells: a modular approach, *Simulation* Vol. 40, No. 3, pp. 3-12

Moniz, A. B. 2012. Anthropocentric-based robotic and autonomous systems: assessment for new organisational options, in Decker and Gutmann (eds.), *Robo- and Informationethics: Some fundamentals*, Zürich/Berlin, LIT, pp. 123 – 157. [http://ideas.repec.org/p/ieu/wpaper/27.html]

Moniz, A. B. & Paulos, Margarida R. (2008). The globalisation in the clothing sector and its implications for work organisation: a view from the Portuguese case, *IET Working Papers Series* 05/2008, Universidade Nova de Lisboa, IET-Research Centre on Enterprise and Work Innovation, Faculty of Science and Technology.

Morioka, S. Sakakibara (2010), A new cell production assembly system with human–robot cooperation, *CIRP Annals - Manufacturing Technology* 59 (2010) 9–12

Paulos, M. R. & Moniz, A. B. (2008). Fragmentation? The future of work in Europe in a global economy: the WORKS final International Conference debate, *Enterprise and Work Innovation Studies*, vol. 4(4), pages 167-169, November.

Ribeiro, L.; Barata, J. and Barreira, P. (2009): Is Ambient Intelligence a truly  Human-Centric Paradigm in Industry? Current Research and Application Scenario, *Enterprise and Work Innovation Studies*, 5, IET, pp. 25 - 35.

Sampaio, J. and Moniz, A. B. 2007. Assessing Human and Technological Dimensions in Virtual Team's Operational Competences, *MPRA Paper* 6942.

Soete, L. 2001. ICTs knowledge work and employment: the challenges to Europe, *International Labour Review*, 140, 2

Thrun, S. (2004). Toward a Framework for Human-Robot Interaction, *Human–Computer Interaction*, 19:1-2, pp. 9-24

Weiss, A., Wurhofer, D., Lankes, M., Tscheligi, M. (2009) Autonomous vs. tele-operated: How people perceive human-robot collaboration with HRP-2. In HRI '09: *Proceedings of the 4th ACM/IEEE International Conference on Human Robot Interaction,* New York, ACM, pp. 257–258.

Williams and Harrison (1999), *Requirements for prototyping the behaviour of virtual environment*s, Human-Computer Interaction Group, Department of Computer Science, University of York.

Yee, N.; Bailenson, J. N.; and Rickertsen, K. (2007). A meta-analysis of the impact of the inclusion and realism of human-like faces on user experiences in interfaces. In CHI '07: *Proceedings of the SIGCHI Conference on Human Factors in Computing Systems,* San Jose, Ca. ACM, New York